\documentclass[letterpaper]{article}

\usepackage{natbib,alifeconf}
\usepackage{todonotes}
\usepackage{color}

\title{Being a leader or being the leader: The evolution of institutionalised hierarchy}

\author{Cedric Perret$^{1}$, Emma Hart$^{1}$ \and Simon T. Powers$^{1}$ \\
\mbox{}\\
$^1$Edinburgh Napier University \\
C.perret@napier.ac.uk} 

\begin{document}
\maketitle

\begin{abstract}
    Human social hierarchy has the unique characteristic of existing in two forms. Firstly, as an informal hierarchy where leaders and followers are implicitly defined by their personal characteristics,  and secondly, as an institutional hierarchy where leaders and followers are explicitly appointed by group decision. Although both forms can reduce the time spent in organising collective tasks, institutional hierarchy imposes additional costs. It is therefore natural to question why it emerges at all. The key difference lies in the fact that institutions can create hierarchy with only a single leader, which is unlikely to occur in unregulated informal hierarchy. To investigate if this difference can affect group decision-making and explain the evolution of institutional hierarchy, we first build an opinion-formation model that simulates group decision making. We show that in comparison to informal hierarchy, a single-leader hierarchy reduces (i) the time a group spends to reach consensus, (ii) the variation in consensus time, and (iii) the rate of increase in consensus time as group size increases. We then use this model to simulate the cost of organising a collective action which produces resources, and integrate this into an evolutionary model where individuals can choose between informal or institutional hierarchy. Our results demonstrate that groups evolve preferences towards institutional hierarchy, despite the cost of creating an institution, as it provides a greater organisational advantage which is less affected by group size and inequality. 
\end{abstract}

\section{Introduction}
Why do humans choose their leaders? A meta-analysis of sixty independent studies shows that leadership effectiveness is not always correlated with leadership emergence \citep{Judge2002PersonalityReview}. In other words, groups sometimes choose incompetent individuals as leaders. For instance, experiments on leader choice showed that \textit{``evaluations of beauty explain success in real elections better than evaluations of competence, intelligence, likability, or trustworthiness''} \citep{Berggren2010TheSuccess}. Yet, despite these risks, most modern human hierarchies spend time and resources to explicitly choose leaders, even if efficient leaders are already designated by their characteristics and skills.

Social organisation plays an important role in the numerous decisions that groups take to efficiently coordinate \citep{Calvert1992LeadershipCoordination}. In social hierarchies, only a minority of individuals (\textit{leaders}) are involved in the decision-making process, while the majority of individuals (\textit{followers}) have limited influence on collective decisions. At the opposite extreme, ancient human hunter-gatherer societies were marked by a relatively equal input from all individuals in group decisions \citep{Boehm2001HierarchyBehavior}. The transition between these two extremes is believed to have been initiated by the advent of agriculture, which created a surplus of resources and increased group size \citep{Bocquet-Appel2011WhenTransition}. In return, larger groups produced more resources thanks to division of labour and specialisation \citep{Pindyck2001Microeconomics}. On the flip side, the need for greater numbers of individuals to coordinate their actions is translated into higher costs of organisation \citep{Calvert1992LeadershipCoordination}. Hierarchy appears as an adaptation to reduce these costs of organisation \citep{VanVugt2011NaturallyLeadership}, and in particular, to address the increase in cost of organisation as a group grows, i.e. \textit{scalar stress} \citep{Johnson1982OrganizationalStress}. In large societies, the benefits created by hierarchy counterbalance the cost of any resulting inequality, eventually leading to its stable emergence.

Human adaptation to hierarchy appeared under two forms \citep{Pielstick2000FormalAnalysis}, expressed in (i) human behaviours \citep{Judge2002PersonalityReview}, and (ii) human preferences. In an \textit{informal} hierarchy, leaders and followers are defined by their intrinsic characteristics. For instance, leader effectiveness is highly correlated with particular psychological traits such as openness and extroversion \citep{Judge2002PersonalityReview}. The second form is \textit{formal} hierarchy where leaders and followers are appointed by group decision. For example, groups confronted by other groups in collective games explicitly elect and identify an individual as a leader \citep{Sherif1954IntergroupExperiment}. We call this form here \textit{institutional} hierarchy to stress that it is supported by institutional rules, which are created by group decision and actively enforced by monitoring and punishment \citep{Ostrom1990GoverningCommons,Hurwicz1996InstitutionsForms}. The emergence of informal or institutional hierarchy can both be explained by the fact that they reduce costs of organisation \citep{Powers2014AnDespotism,Perret2017EmergenceModel}. However, institutional hierarchies are surprisingly pervasive in modern societies, given that they carry additional costs in comparison to informal ones. A key to this puzzle lies in the particularity of institutions which allow humans to hand-tune their behaviours, e.g. by designating a single leader, in comparison to informal hierarchies in which leaders emerge through blind evolutionary processes. However, it remains unclear whether this difference could drive the appearance of institutional hierarchies. 

Currently, independent explanations for the evolution of informal and institutional hierarchy have been provided \citep{Powers2014AnDespotism,Perret2017EmergenceModel}, but there is no model that investigates the competition between these two forms of social organisation. To fill this gap, we first investigate if the single-leader model found in institutional hierarchy facilitates group decision-making, i.e. leads to shorter coordination times. Second, we evaluate whether this benefit is sufficient to lead to the evolution of cultural preferences toward institutional hierarchy despite the additional costs of maintaining the institution.
To do so, we describe group decision-making using an opinion-formation model
\citep{Castellano2009StatisticalDynamics} that simulates a sequence of discussions between individuals, and has been shown to reflect the organisational advantage brought by leaders \citep{Gavrilets2016ConvergenceLeadership}. We define leaders and followers by their capacity to influence others, and analyse the effect of the number of leaders on the time a group spends to reach consensus. We then integrate this model into an evolutionary model where the time spent to reach consensus is translated into the cost of organising collective tasks. The model simulates a population structured around patches where individuals organise and carry out a collective action, which produces additional resources. Individuals can choose between informal social organisation where leaders and followers are defined by individuals' characteristics, or institutional social organisation where leaders and followers are defined by the institution. Our results show that in comparison to informal hierarchy, hierarchy with a single leader reduces (i) the consensus time, (ii) the variation in the consensus time, and (iii) the increase in consensus time as group size increases. We demonstrate that individuals evolve cultural preferences towards institutional hierarchy because it provides a greater organisational advantage than informal hierarchy, and reduces the detrimental effect of group size and inequality on the time spent to organise collective actions.

\section{The effect of the number of leaders on group decision-making}
We define social organisation by the proportion of leaders and followers present in a patch. This ranges from a perfect egalitarian organisation described by all individuals being followers or leaders, to the most hierarchical organisation with one leader and the rest of the group as followers. We define political organisation as the process by which leaders and followers are defined. The political organisation of a group can either be \textit{informal}, i.e. leaders and followers are defined by default by individual characteristics, or \textit{institutional}, i.e. leaders and followers are defined by group decision \citep{Hurwicz1996InstitutionsForms}. It is worth noting that we constrain an institutional group to be a hierarchy, but a group can have an informal political organisation with either an egalitarian or hierarchical social organisation.

\subsection{Model definition}
We develop an opinion-formation model to simulate group decision-making based on previous work \citep{Deffuant2000MixingAgents,Perret2017EmergenceModel}. It is an individual-based model which consists of a sequence of discussions between individuals until their opinions are close enough i.e. the group has reached a global consensus. Opinion-formation models are well-known tools to study social dynamics \citep{Castellano2009StatisticalDynamics}, and have been shown to reflect the benefit of leaders on group decision-making \citep{Gavrilets2016ConvergenceLeadership,Perret2017EmergenceModel}. Individuals are described by an opinion $x$, and a value of influence $\alpha$. These are both continuous values defined on [0,1]. The trait $\alpha$ represents the influence of an individual and affects (i) the capacity of one individual to modify the opinion of another individual towards its own opinion, (ii) the reluctance of an individual to change its opinion, and (iii) the probability that an individual talks to other individuals. These three traits, i.e. persuasiveness, stubbornness and talkativeness, are highly correlated in leaders personalities \citep{Judge2002PersonalityReview} and are the key factors in explaining how leaders reduce time to reach consensus \citep{Gavrilets2016ConvergenceLeadership}. Individuals can have one of two profiles: a leader $l=1$ with a high influence value $\alpha_\mathrm{l}$, or a follower $l=0$ with a low influence value $\alpha_\mathrm{f}$, where $\alpha_\mathrm{l} > \alpha_\mathrm{f}$. 

The opinion $x$ is randomly generated at the beginning of the opinion formation. At each time-step, there is a discussion event where one speaker talks to $N_\mathrm{l}$ listeners to bring the followers' opinion closer to its own. The probability $P$ of an individual $i$ to be chosen as a speaker $\mathrm{u}$ is an increasing function of its $\alpha$ value as follows:
\begin{equation}
P_{i}(t)=\frac{(\alpha_{i}(t))^k}{{\sum_{n=1}^{N}(\alpha_{n}(t))^k}}.
\label{eqn:speakerProbability}
\end{equation}
In the simulations we chose $k = 4$  so that in a group of large size i.e. $1000$ individuals, with the most extreme hierarchy (one leader with maximum influence, $N-1$ followers with minimum influence), the probability that a leader is chosen as a speaker is close to $90\%$. 
The speaker talks with $N_\mathrm{l}$ listeners randomly sampled within the other individuals in the group. This limit on the number of listeners models time constraints, and cognitive constraints of human brains \citep{Dunbar1992NeocortexPrimates}. We assume that every individual can be chosen as a listener, i.e. the social network is a complete network, in order to avoid explicitly modelling the network structure and to keep the model tractable. We also consider that individuals interactions are not limited to individuals with close opinions i.e. bounded confidence, because this model describes a consensus seeking process where individuals are willing to convince each other. During a discussion event, a listener $\mathrm{v}$ updates its preference to a value $x'_\mathrm{v}$ following the equation below, where $\mathrm{v}$ represents the listener and $\mathrm{u}$ the speaker:
\begin{equation}
x'_\mathrm{v} = x_\mathrm{v} + (\alpha_\mathrm{u}-\alpha_\mathrm{v})(x_\mathrm{u} - x_\mathrm{v}).
\label{eqn:opinionUpdate}
\end{equation}
We assume that the position of speaker gives a slight influential advantage over the listeners. Therefore, the minimum difference of influence $\alpha_\mathrm{u}-\alpha_\mathrm{v}$ is set to a positive low value, here $0.01$. This assumption is necessary to avoid a systematic convergence of the preferences towards the individual with the highest $\alpha$, a phenomenon not observed in real life. The individuals repeat the previous step until consensus is reached, i.e. the standard deviation of the preferences $x$ is less than a threshold $x_\mathrm{\theta}$. The number of discussion events that occurred to reach consensus is called the consensus time, $t^*$. 

\subsection{Analysis}

We use the opinion-formation model to investigate the difference in consensus time between hierarchy with a single leader and multiple leaders. Because of this heterogeneity, we use numerical simulations to analyse the model. The default parameters are for the consensus threshold $f_{\mathrm{\theta}} = 0.05$, the number of listeners $N_\mathrm{l} = 50$, the influence of leaders $\alpha_\mathrm{l} = 0.75$ and the influence of followers $\alpha_\mathrm{f} = 0.25$. The results presented are the mean across $1000$ replicates. The error bars represent the standard error from the mean.

\begin{figure}[t]
\begin{center}
\includegraphics[width=.8\linewidth]{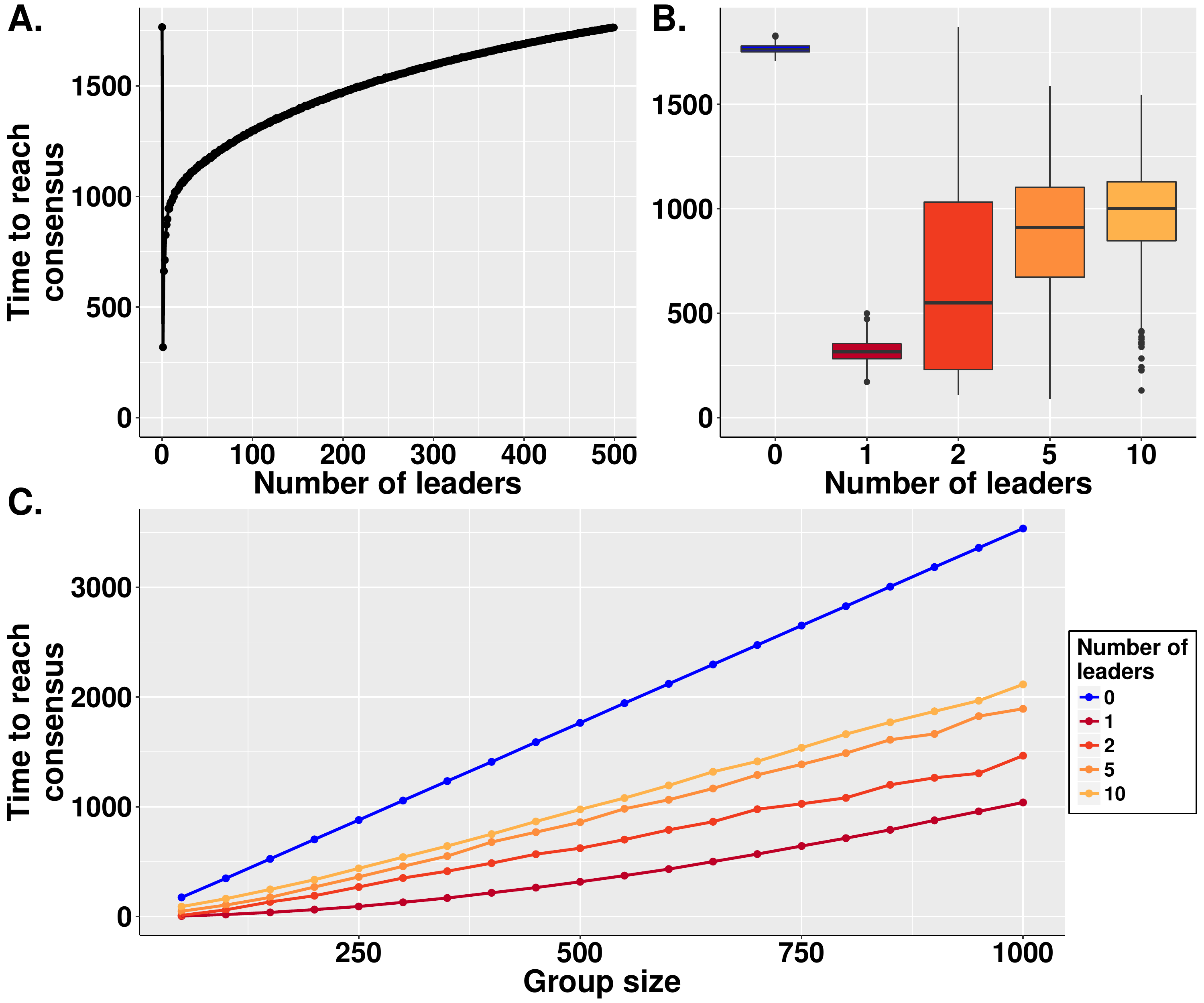}
\caption{Effect of number of leaders on decision-making. (A,B) Mean consensus time $t^*$ as a function of number of leaders in a group of $500$ individuals (C.) Mean consensus time $t^*$ as a function of number of leaders and group size.}
\label{fig:opi_L_LN}
\end{center}
\end{figure}

Figure \ref{fig:opi_L_LN}.A confirms that hierarchy (i.e. a small  number of leaders) provides an organisational advantage by reducing the consensus time. Figure \ref{fig:opi_L_LN}.B shows that (i) the presence of a single leader reduces the average consensus time compared to multiple leaders, and (ii) the presence of a single leader assures a consistently lower consensus time (shown by the low variance). The presence of two leaders provides a variable advantage, which ranges from the same result as the single leader to the result from a group without a leader. As the number of leaders increases, the time to consensus increases while the variability decreases. Finally, Figure \ref{fig:opi_L_LN}.C shows that the rate of increase of consensus time grows more slowly with group size when the number of leaders is smaller. In other words, the benefit of single leader increases with group size.

\begin{figure}[t]
\begin{center}
\includegraphics[width=.8\linewidth]{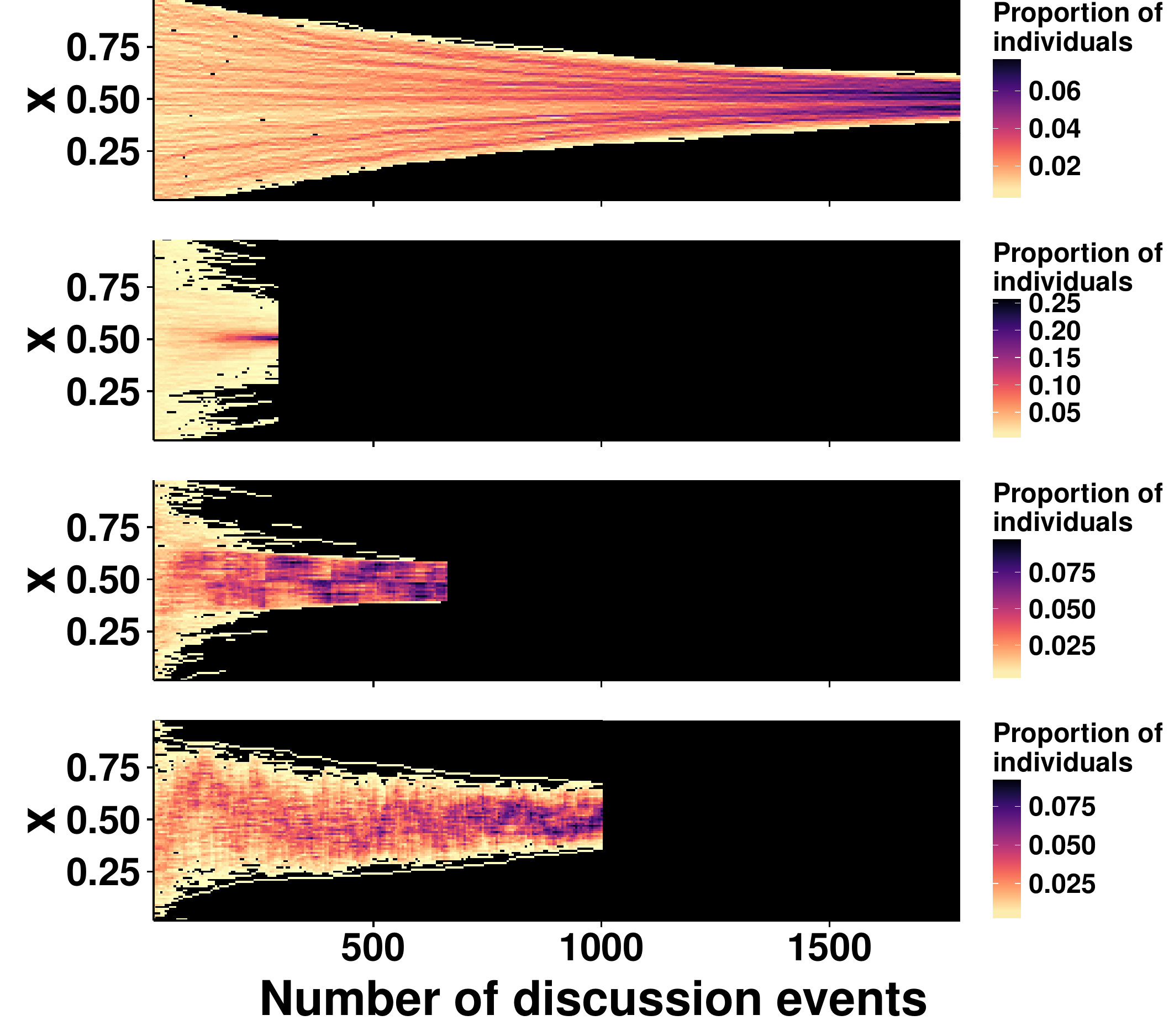}
\caption{Density distribution of individual opinion as a function of number of discussion events for different number of leaders: from top to bottom $0, 1, 2, 10$. For illustration, the difference between the opinions of leaders are set to be maximum and equidistant.}
\label{fig:x_propL}
\end{center}
\end{figure}
Figure \ref{fig:x_propL} illustrates the opinion formation and the effect of the number of leaders on group decision-making. First, we see that in the absence of leaders, or with a single leader, individuals' opinions slowly and consistently converge. The presence of a single leader speeds up this process as the leader quickly convinces the majority of the group. The presence of multiple leaders creates a more heterogeneous pattern of convergence. The presence of two leaders results in the majority switching from one leader to another: leaders alternatively convince individuals from the group but neither leader has enough followers to reach consensus. When more than two leaders are present, the majority of opinion fluctuates between the different leaders. In both cases, leaders' stubbornness slows convergence of leaders towards the others, which in turn slows down the whole process. To conclude, hierarchy with a single leader clearly provides a benefit to group organisation which is (i) stronger, (ii) more constant, and (iii) more resistant to group size increase than multiple leaders. Thus, a slight change in the number of leaders can have a drastic effect on group organisation.

\section{An evolutionary model of political organisation}
We now develop an evolutionary model to investigate if the benefit of single leader hierarchy is sufficient to lead to the evolution of cultural preferences towards institutional hierarchy. Individuals carry two evolving traits: their social personality $s$ and their preference for political organisation $h$. The trait $s$ represents the intrinsic personality of an individual in a social interaction (e.g. talkativeness, boldness, charisma) and can be either dominant $s=1$, or compliant $s=0$. It defines an individual's influence $\alpha$ in informal organisation, and the probability to be chosen as a leader in institutional organisation. The trait $h$ represents the preference in terms of political organisation of an individual: $0$ represents a preference for informal organisation, and $1$ a preference for institutional organisation. In addition, individuals are described by a value of influence $\alpha$ as described previously. The influence is either defined by an individual social personality $s$ in an informal hierarchy, or by their assigned individual social position in institutional hierarchy (explained below). The initial values of the social personality of individuals, $s$, are randomly generated. The initial values of preference for political organisation $h$ are set to $0$ to represent the initial absence of institutions. The two traits ${s, h}$ carried by individuals are transmitted vertically from parent to offspring, e.g. by social learning as is common in hunter-gatherer groups \citep{Hewlett2011SocialHunter-gatherers}. They mutate following a mutation rate of $\mu$. As these traits are assumed to be at least partly cultural, the mutation rate is higher than for a classical genetic trait. When a mutation occurs, the trait value is flipped.

\subsubsection{Life cycle and social traits}
We consider an island model with a population of individuals that is subdivided into a finite number of patches $N_\mathrm{p}$ \citep{Wright1931EvolutionPopulations}. The life cycle consists of discrete and non-overlapping generations as follows:
\begin{enumerate}
    \item Individuals decide whether to create an institutional hierarchy and appoint a leader; or defaults to an informal organisation where leaders and followers roles are defined by individuals' personality $s$.  Individuals creating an institutional hierarchy pay a cost $c_\mathrm{h}$.
    \item Individuals play a decision-making game on their patch as defined above (equations \ref{eqn:speakerProbability}, \ref{eqn:opinionUpdate}). The time taken to reach consensus is translated into an opportunity cost of organisation (equation \ref{eqn:costOrga}).
    \item After consensus is reached, all individuals on a patch take part in a collective task which produces an amount of extra resource, discounted by the cost of organisation (equation \ref{eqn:benefit}). 
    \item The resource obtained from the collective task is distributed among all individuals on the patch. Leaders get a surplus of resources modulated by a parameter $d$ which modulates the inequality between leaders and followers (equation \ref{eqn:distribution})
    \item Individuals produce a number of offspring drawn from a Poisson distribution, with the mean determined by the resources received (equation \ref{eqn:fitness})
    \item All individuals of the previous generation perish.
    \item Offspring migrate with a fixed probability $m$. Migrating individuals enter a patch chosen at random from the population (excluding their natal patch).
\end{enumerate}

\subsubsection{Political organisation}
Each group within a patch is defined by a political organisation $h^*$. At the beginning of each generation, individuals decide if they want to design an institutional hierarchy and appoint a leader ($h^* = 1$); this occurs if the majority of individuals in the group have a preference toward institutional hierarchy i.e. $\frac{1}{N_j(t)}\sum_i^{N_j(t)}h_{ij}(t) > 0.5$. In the absence of institutions ($h^* = 0$), a group is organised by default as an informal hierarchy.

In an institutional hierarchy, one single leader is randomly selected from the individuals with dominant personality $s=1$ and its influence is set to $\alpha_\mathrm{l}$. The rest of the individuals within the patch adopt a follower profile and their influences are set to $\alpha_\mathrm{f}$ (independently of their social personality). In an informal hierarchy, an individual's influence $\alpha$ is defined by its social personality with $\alpha_\mathrm{l}$ for dominant individuals $s = 1$ and $\alpha_\mathrm{f}$ for compliant individuals $s = 0$. In order to be sustainable, institutions require resources to monitor individuals and punish transgressors \citep{Ostrom1990GoverningCommons}. Thus, individuals creating an institutional hierarchy pay a cost $c_\mathrm{h}$. 

\subsubsection{Organisation by decision-making}
Once individuals have chosen their political organisation, they organise a collective task through group decision-making as described above. The consensus time is translated into a cost of organisation:
\begin{equation}
C_\mathrm{oj}(t) = t_j^* C_t
\label{eqn:costOrga}
\end{equation}
The cost of organisation comes from the time dedicated to organisation instead of carrying out the actual task -- groups that take too long to reach a decision may lose resources or pay other opportunity costs. This cost is modulated by $C_\mathrm{t}$, which is a parameter representing the time constraint on decision making and depends of the limitation of time on the task, for instance, the speed of depletion of resources or the need to build defences before an enemy arrives. We consider here that the final decision reached has no effect on the benefit produced by the collective task -- the benefit is only affected by the time taken to reach consensus. 
\subsubsection{Collective task}
At each generation, individuals take part in a collective task and produce additional resources $B_j(t)$:
\begin{equation}
B_j(t) = \frac{\beta_\mathrm{b}}{1+e^{-\gamma_\mathrm{b} (N_j(t)-b_{\mathrm{mid}})}} - C_\mathrm{oj}(t).
\label{eqn:benefit}
\end{equation}
The collective task simulates the numerous cooperative tasks realised during the lifetime of an individual. It can encompass many actions such as hunting of large game or construction of an irrigation system. The benefit is calculated from a sigmoid function described by $\beta_\mathrm{b}$, $b_{\mathrm{mid}}$ and $\gamma_\mathrm{b}$, respectively the maximum, the group size at the sigmoid's midpoint, and the steepness of the increase in the benefit induced by additional participants. We assume economy of scale in which additional participants increase the benefit super-linearly \citep{Pindyck2001Microeconomics}. But as is standard in micro-economic theory, we also make the conservative assumption that the benefit of the collective task eventually has diminishing marginal returns which overcomes the economy of scale because of other limiting factors \citep{Foster2004DiminishingCommons}. 

\subsubsection{Distribution of resources}
The resources produced by the collective task are distributed between the individuals on a given patch. The share of an individual, $p_{ij}(t)$, is then equal to: 
\begin{equation}
p_{ij}(t) = \frac{1 + l_i(t)d}{\sum_{i=1}^{N_j}(1 + l_i(t)d)}.
\label{eqn:distribution}
\end{equation}
Leaders ($l=1$) receive a surplus of resources modulated by the level of ecological inequality $d$. For $d = 0$, the distribution within a patch is egalitarian and the influence of individuals does not affect the share of each individual. Such a scenario is close to that observed in societies of pre-Neolithic hunter-gatherers. For $d = 1$, leaders receive twice the amount a follower receives. It is assumed for simplicity that $d$ is the same for all patches, and is determined for example by the state of technology, e.g. food storage and military technologies. 

\subsubsection{Reproduction}
After receiving their share of the additional resources, individuals have a number of offspring sampled from a Poisson distribution centred on the individual fitness, $w$. The fitness of individual $i$ on patch $j$ at time $t$ is described by the following equation, where $N_j(t)$ is the total number of individual on patch $j$:
\begin{equation}
w_{ij}(t) = \frac{r_\mathrm{a}}{1+\frac{N_j(t)}{K}} + r_{\mathrm{b}ij}(t) - c_\mathrm{h}h^*_j - c_\mathrm{n}s_{ij}.
\label{eqn:fitness}
\end{equation}
The fitness of an individual is the sum of an intrinsic growth rate $r_\mathrm{a}$ limited by the carrying capacity $K$, and additional growth rate resulting from the extra resources produced by the collective task, $r_{\mathrm{b}ij}(t)$. The fitness of individuals with institutional organisation is discounted by a cost of institution $c_\mathrm{h}$, which represents the cost to monitor and enforce the institutional rule. The fitness of dominant individuals is discounted by a cost of negotiation $c_\mathrm{n}$ which represents the extra time and resources that an individual with dominant personality allocates to persuade others. 
The additional growth rate $r_{\mathrm{b}ij}(t)$ is calculated as follows:
\begin{equation}
r_{\mathrm{b}ij}(t) = \beta_\mathrm{r}(1-e^{- \gamma_\mathrm{r}(B_j(t)p_{ij}(t))}).
\label{eqn:additionalGrowth}
\end{equation}
The term $r_{\mathrm{b}ij}(t)$ is calculated from a logistic function described by $\gamma_\mathrm{r}$ and $\beta_\mathrm{r}$, respectively the form and the maximum of the increase in growth rate induced by the additional resources. The additional resources are given by the total amount of benefit, $B_j(t)$, multiplied by the share the individual receives, $p_{ij}(t)$. The increase of the growth rate follows a logistic relation because of the inevitable presence of other limiting factors.
After reproduction, offspring individuals migrate with a probability equal to a fixed migration rate $m$. Migrating individuals enter a patch chosen at random from the population (excluding their natal patch).

\subsection{Analysis}

We use this model to answer the following question: \textit{Can the organisational benefit of single leader hierarchy lead to a transition from informal to institutional organisation despite the additional cost of institutions?} Because of the non-linearities of the model, which result from the interactions of all of the variables, we analyse it using replicated numerical simulations. We focus on the effect of the following parameters: (i) the level of ecological inequality $d$ (ii) the cost of institution $C_\mathrm{h}$  and (iii) the time constraint $C_\mathrm{t}$.
The default parameters used in the simulations, unless otherwise specified, are $N_p = 50$, $N_j(0) = 20$, $K = 20$, $r_\mathrm{a} = 2$, $\beta_\mathrm{b} = 10000$, $\gamma_\mathrm{b} = 0.005$, $b_{\mathrm{mid}} = 250$, $\beta_\mathrm{r} = 3$, $\gamma_\mathrm{r} = 0.05$, $\mu_\mathrm{m} = 0.01$ and $m = 0.05$. These parameters are chosen in order to allow the transition between tribe size (50 to 100 individuals) to chiefdom size (1000 individuals). The default parameters for the group decision-making are the same as previously. Finally, we want to allow for hierarchy even when the political organisation is informal. To do so, we choose a high cost of negotiation $C_\mathrm{N}$ which limits the evolution of too many leaders and allows relatively stable informal hierarchy. The results presented are the mean across $32$ replicates when the result is as a function of generations; and across $32$ replicates and $5000$ generations when the results are as a function of a parameter. Where the result is described as a mean, it is the mean value across patches. The error bars represent the standard error from the mean and are not represented when they are too small to be visible ($<5\%$ of the maximum value).

\begin{figure}[t]
\begin{center}
\includegraphics[width=.8\linewidth]{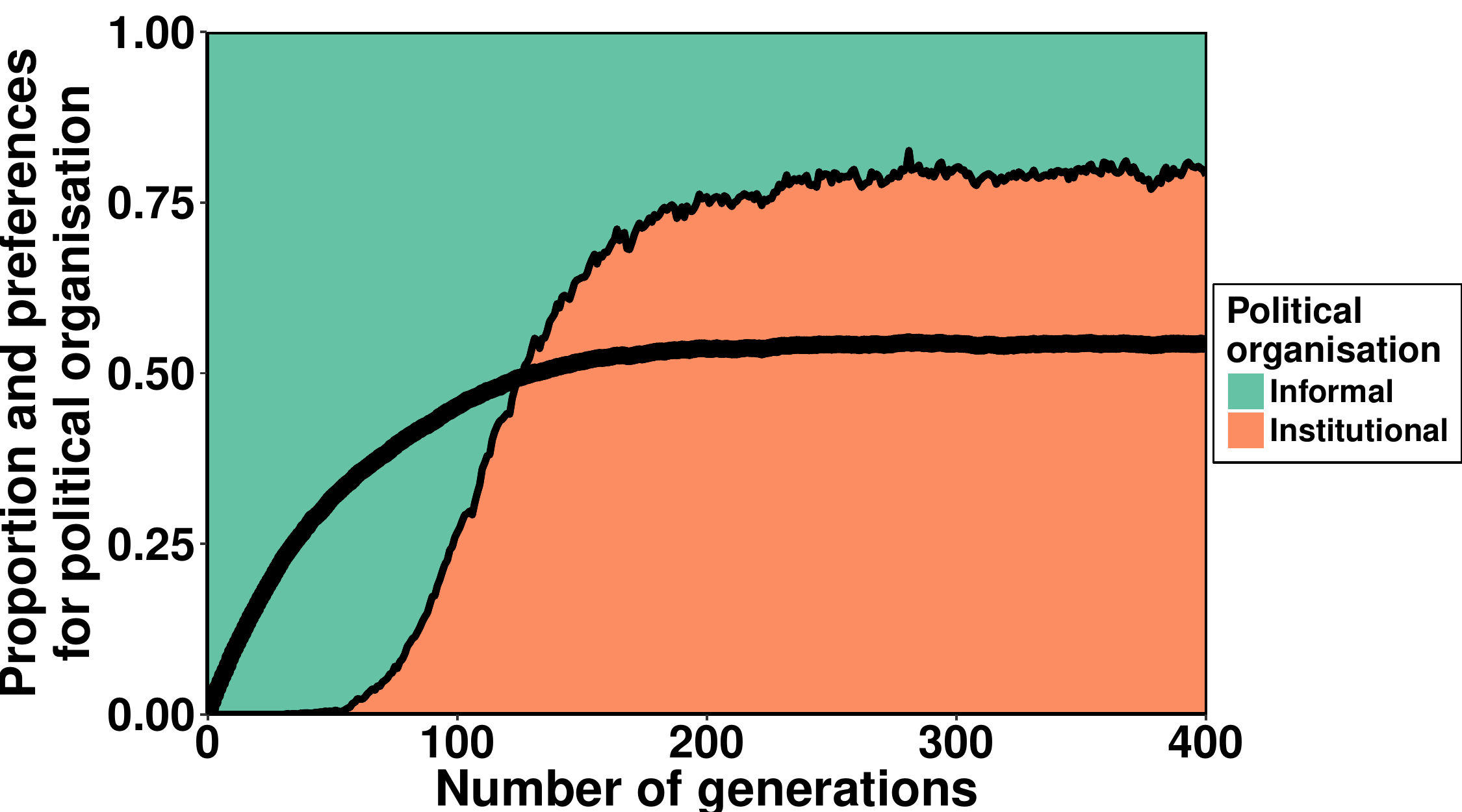}
\caption{Evolution of the distribution of political organisation $h^*$ (colour) and mean proportion of individuals with preferences towards institutional hierarchy across generations.}
\label{fig:orga_gen}
\end{center}
\end{figure}

\begin{figure}[t]
\begin{center}
\includegraphics[width=.8\linewidth]{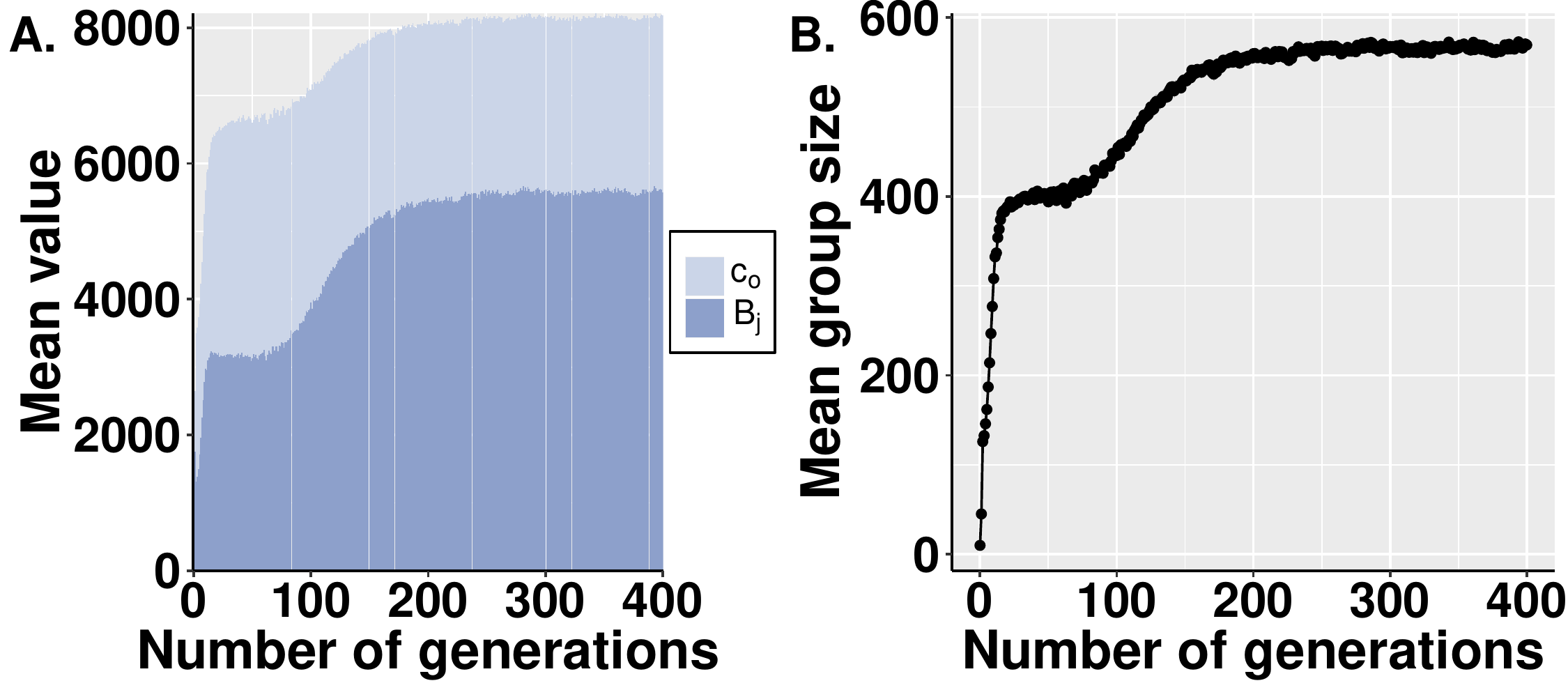}
\caption{(A) Evolution of mean additional resources $B$ (dark) equal to total resources produced discounted by cost of organisation $C_\mathrm{o}$ (light) across generations. (B) Evolution of mean group size across generations.}
\label{fig:b_gen}
\end{center}
\end{figure}
\begin{figure}[t]
\begin{center}
\includegraphics[width=.8\linewidth]{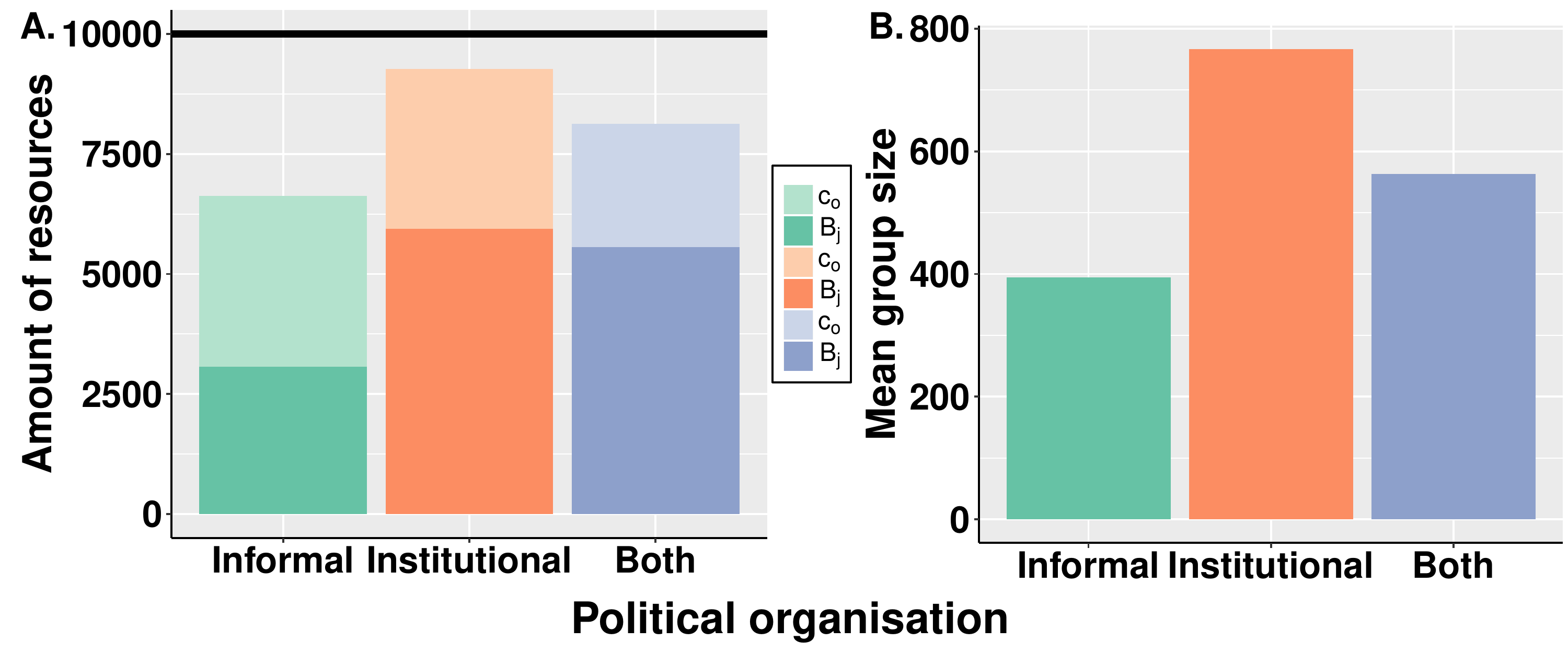}
\caption{(A) mean additional resources $B$ (dark) equal to total resources produced discounted by cost of organisation $C_\mathrm{o}$ (light), and (B) mean group size between simulations where are only allowed either institutional, informal or both organisations.}
\label{fig:B_orga}
\end{center}
\end{figure}

Figure \ref{fig:orga_gen} demonstrates that for a moderate cost of institution, individual preferences evolve towards institutional hierarchy and thus, most of groups switch from informal to institutional hierarchy. Groups have in average only slightly more than $50\%$ of individuals with preference toward institutional hierarchy because having any proportion above $50\%$ has the same effect on political organisation and therefore the fitness of all individuals within the group. The small proportion of groups with informal hierarchy are explained by the cost of the institution and random mutations in individual's preferences, which can lead some groups to temporarily switch back to informal hierarchy. The prevalence of institutional hierarchy remains stable for long period ($5000$ generations). Figure \ref{fig:b_gen} shows that the total amount of resources produced and thus the group size increases through time. The cost of organisation also increases but remains low enough so that a large group provides more resources than a small group. Figure \ref{fig:b_gen} shows that two increases in production and group size happen. The first corresponds to the emergence of informal hierarchy, and the second to the subsequent emergence of institutional hierarchy. This result and the results presented in Figure \ref{fig:B_orga} demonstrate that institutional hierarchy allows a higher production and a larger group size.  This is because a group with institutional hierarchy has (i) a lower cost of organisation and, (ii) a larger production of surplus resources due to the larger size they reach. When both types of organisation are allowed, groups reach an intermediate size and productivity because of the cost of institution and the presence of a minority of small groups with informal hierarchy. To summarise, groups developing institutional hierarchy strongly reduce their cost of organisation. They grow larger, which improves their productivity, while hierarchy limits the increase in the cost of organisation. As a consequence, these groups export a greater number of migrants, who carry their cultural preferences for institutions to other groups, leading to the global spread of institutions.

\begin{figure}[t]
\begin{center}
\includegraphics[width=.8\linewidth]{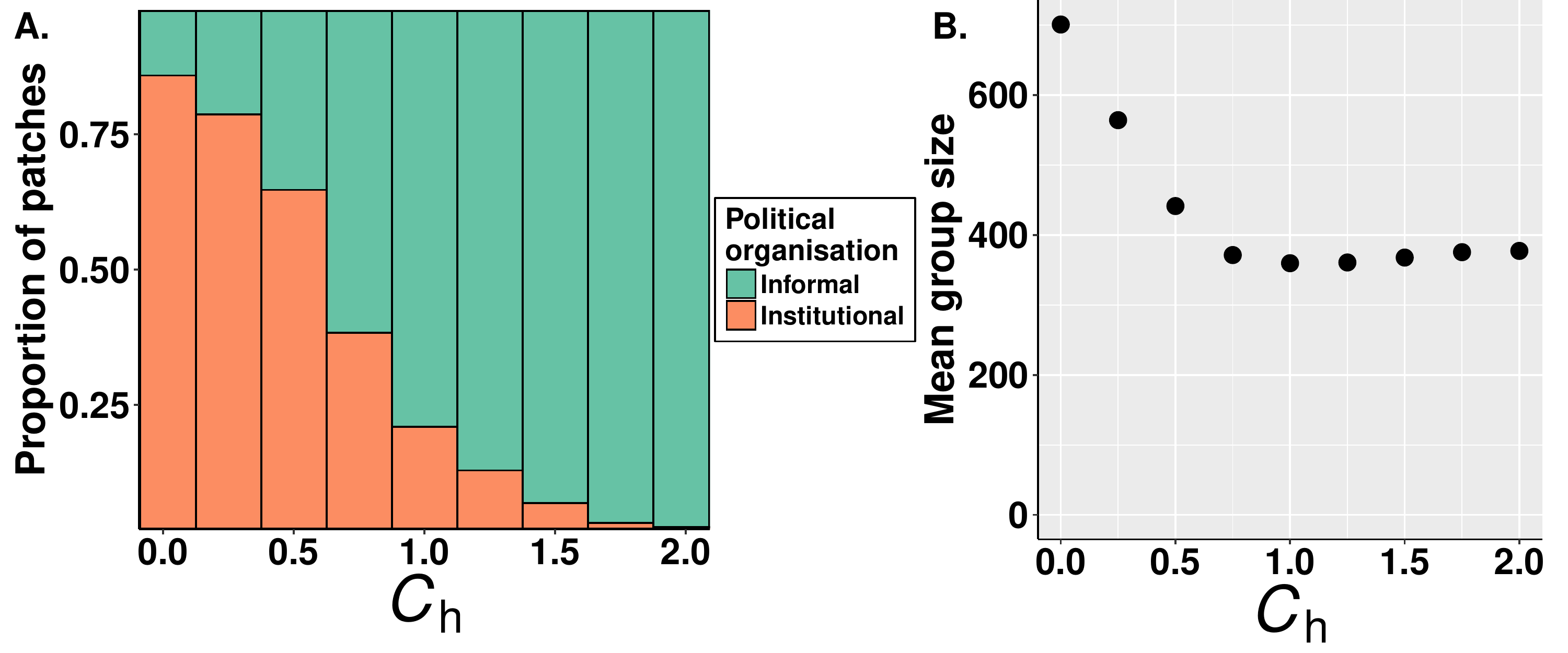}
\caption{Distribution of (A) political organisation $h^*$ and (B) mean group size as a function of the cost of an institution $C_\mathrm{h}$.}
\label{fig:orga_CH}
\end{center}
\end{figure}

\begin{figure}[t]
\begin{center}
\includegraphics[width=.8\linewidth]{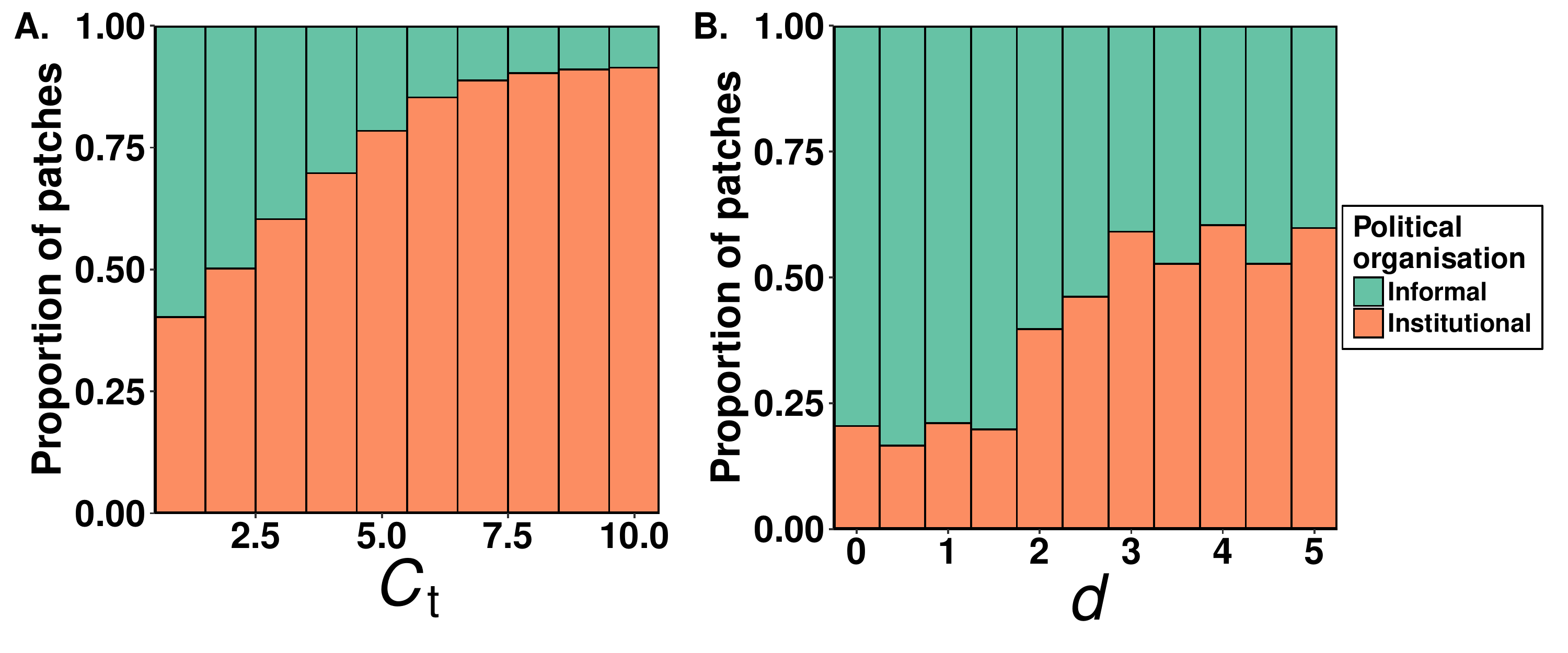}
\caption{(A) Distribution of political organisation $h^*$ as a function of time constraint $C_\mathrm{t}$. (B) Distribution of political organisation $h^*$ as a function of level of ecological inequality $d$ with $C_\mathrm{h}=1$}.
\label{fig:orga_CT_d}
\end{center}
\end{figure}

Figure \ref{fig:orga_CH} shows that an increase in the cost of institution $C_\mathrm{h}$ reduces the proportion of institutional hierarchy and the average group size. This result is explained by the high cost of institution overcoming the benefit brought by institutional hierarchy. However, institutional hierarchy still evolves even for a moderate cost of institutions. Indeed, a cost of $1$ means that all individuals within a group need a growth rate twice higher and thus, to produce approximately twice as much resources to sustain the same fitness (see equation \ref{eqn:fitness}. Moreover, Figure \ref{fig:orga_CH} shows that individuals develop institutional hierarchy even if it doesn't significantly modify the average group size e.g. same size between $C_\mathrm{h}=1$ and $C_\mathrm{h}=2$. This is explained by single leader hierarchy providing a more constant organisational benefit than the multiple leaders of informal hierarchy. 
Figure \ref{fig:orga_CT_d}.A shows that a larger proportion of groups develop institutional hierarchy when the time constraint on the decision making $C_\mathrm{t}$ is high e.g. a time limited task such as warfare. This is because the shorter consensus time brought by single leader hierarchy has more consequences on the absolute group production.

\begin{figure}[t]
\begin{center}
\includegraphics[width=.8\linewidth]{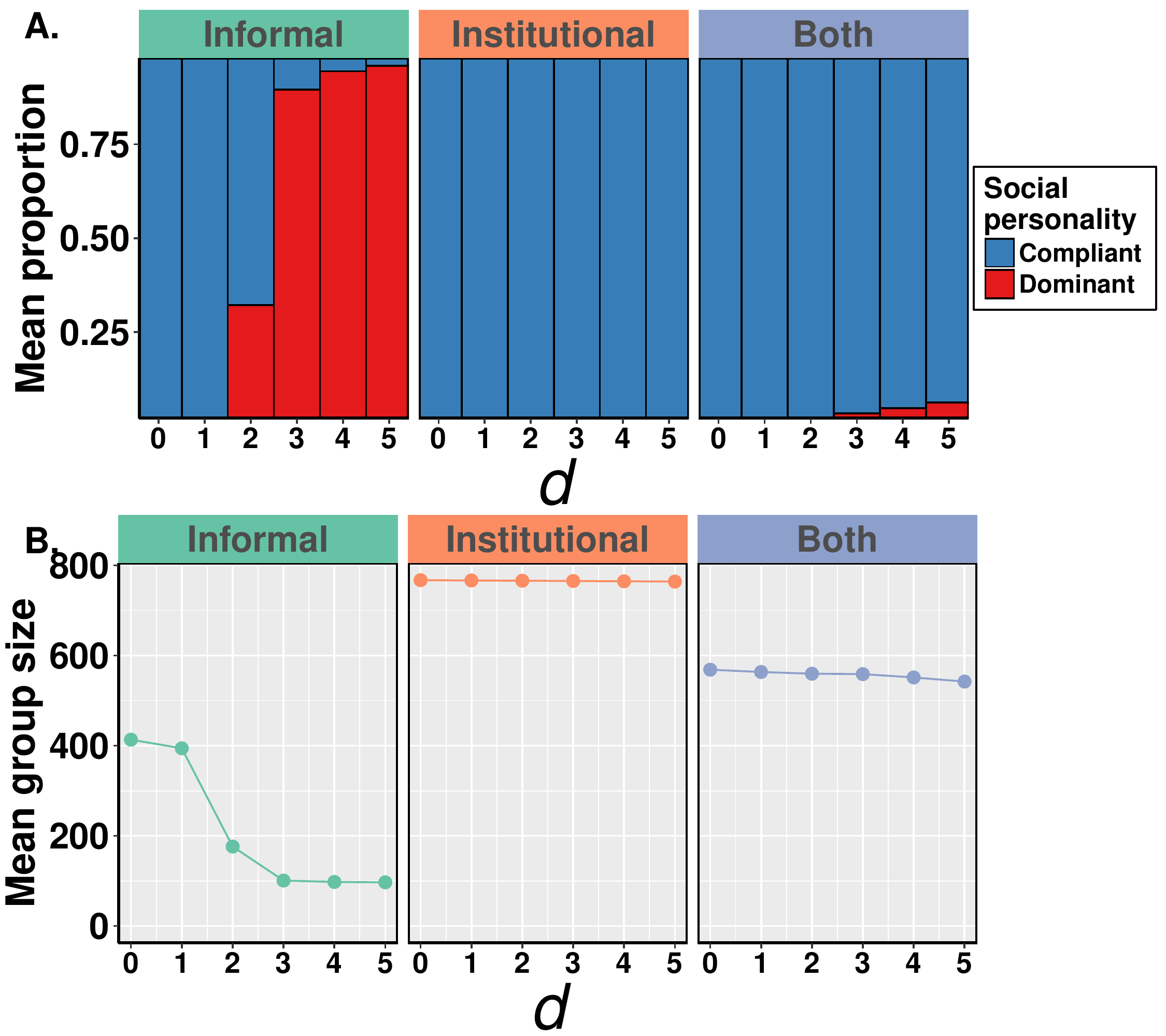}
\caption{(A) Mean distribution of social personality and (B) mean group size as a function of the level of ecological inequality $d$.}
\label{fig:d__sProp_size}
\end{center}
\end{figure}

Figure \ref{fig:orga_CT_d}.B shows that a higher proportion of groups develop institutional hierarchy when the level of ecological inequality $d$ is higher. This result is explained by Figure \ref{fig:d__sProp_size} which shows that the benefit provided by institutional hierarchy persists even under high inequality. On the contrary, Figure \ref{fig:d__sProp_size}.A shows that in an informal organisation, an increase in the level of inequality leads to an increase in the number of leaders. This results in a collapse of hierarchy, a high cost of organisation and smaller group size (Figure \ref{fig:d__sProp_size}.B). This difference in the effect of inequality is explained by institutional hierarchy having only one expressed leader even if multiple individuals want to be leaders. In addition, only one individual attains the status of leader and hence receives a surplus of resources, which ultimately limits the increase in number of dominant individuals.

\section{Discussion}

Human social hierarchy can be formed because individuals act as leaders and followers, i.e. informal hierarchy, or because certain individuals are chosen as leaders and followers, i.e. institutional hierarchy. But why do human groups create costly institutional hierarchies if hierarchy already emerges naturally from individual behaviours? The key difference is that single leaders can appear in institutional hierarchy designed by group decision, but are highly unlikely in informal organisation shaped by blind evolution of personality traits. Thus, in this paper, we have focused on the difference between single and multiple leader hierarchies and have shown that institutional hierarchy with a single leader reduces more (i) the consensus time, (ii) the variation in the consensus time, and (iii) the increase in consensus time as a group grows. Our evolutionary model demonstrates that this difference results in individuals' preferences evolving towards institutional hierarchy even if this has an additional cost. To conclude, group organisation is facilitated by hierarchy but is highly intolerant to multiple leaders. This particularity provides one possible explanation for the evolution and wide spread of institutional hierarchy. To understand how critical and general is this explanation, further work should (i) explore more widely the model and its parameters and (ii) use data to test the prediction e.g. compare the cost of organisation in informal and institutional hierarchy. 

The results of the opinion-formation model confirm previous work which shows that an informal leader with the features defined here speeds up consensus time \citep{Gavrilets2016ConvergenceLeadership}. This prior work showed that an increase in the number of leaders slows down the consensus, because it creates more stubborn individuals. Our result adds that multiple leaders also slow down the consensus, because leaders persuade each others' followers, creating conflict of interest between a large proportion of the group. It results in a more detrimental effect of multiple leaders on consensus time, which is amplified by group size. Previous theoretical work have investigated the emergence of either informal or institutional hierarchy, but ignored the competition between the two forms. \citet{Powers2014AnDespotism} developed an evolutionary model in which individuals favour institutional hierarchy over an egalitarian organisation. Other theoretical models have shown that a similar process can drive the evolution of individuals towards leader and follower behaviours, thus creating an informal hierarchy \citep{Johnstone2011EvolutionLeadership,Perret2017EmergenceModel}. We confirm and connect these works by showing that institutional hierarchy can be favoured over informal hierarchy because it provides additional benefit to group decision-making, in terms of consensus time.

Our model predicts that institutional hierarchy evolves when (i) group size is high (and so productivity and cost of organisation are high), and (ii) inequality is high. These predictions fit with the environmental and social changes observed following the advent of agriculture. Agriculture created a durable surplus of resources which increased productivity and inequality \citep{Bocquet-Appel2011WhenTransition,Mattison2016TheInequality}. However, our model also predicts that the productivity benefit of institutional hierarchies can be counterbalanced by a high cost of institutions. It is hard to evaluate the costs implied by institutions, but it is worth noting that they result mostly from the resources and time allocated to monitor and punish individuals not complying with the rules, i.e. here individuals trying to become leaders. Our model has shown that institutional hierarchy limits the number of individuals aspiring to become leaders, and thus suggests that the costs of institutions remain limited even in large groups. 
It is worth noting that instead of competing, the two forms of political organisation could have interacted and even facilitated the development of each other. First, the development of informal hierarchy also leads to a higher group size and higher inequality.
Second, the influence of an individual is in truth defined by both an individual's personality and its social position. Integrating a composite value of influence in this model could provides more insight into the interactions between these two forms of political organisation.

In this model, we have explored only one form of institution and one function of hierarchy. It would be interesting to explore other types of institutions, such as those allowing multiple levels of hierarchy, or restrict the number of people involved in the decision-making, as found in representative democracy. Other functions of hierarchy could also be investigated, e.g. to enforce cooperation \citep{Hooper2010AGroups}. However, it is worth noting that extending the model to integrate the possibility of voting for more leaders would carry similar qualitative results with individuals evolving a preference toward one leader. The presence of multiples leaders appears only later in human history, with the rise of complex states composed of multiple layers of hierarchy that constrain the behaviour of different leaders \citep{Johnson2000TheState}.

Institutions are believed to be crucial innovations for the emergence of human societies. We have shown here that one of their major benefit is to provide humans with a finer tool to modify their behaviour, which can be crucial for some processes such as shown here with hierarchy. More than a new innovation, the development of institutions marks a transition in the dynamics shaping human behaviours: from long and blind evolutionary process to fast cultural dynamics. 

\footnotesize
\bibliographystyle{apalike}
\bibliography{Mendeley.bib}

\end{document}